\documentclass[final,5p,times,twocolumn,sort&compress]{elsarticle}
\usepackage{amssymb}
\usepackage{url}
\usepackage{color}
\usepackage{gensymb}
\usepackage{textcomp}
\usepackage{cmap}
\journal{Ultramicroscopy}

\begin{document}


\begin{frontmatter}

\title{Nanometer scale elemental analysis in the helium ion microscope using time of flight spectrometry}

\author[hzdraddress,tudaddress]{N. Klingner\corref{mycorrespondingauthor}}
\cortext[mycorrespondingauthor]{Corresponding author}
\ead{n.klingner@hzdr.de}

\author[hzdraddress]{R. Heller}
\author[hzdraddress]{G. Hlawacek}
\author[hzdraddress]{J. von Borany}

\author[Zeissaddress]{J. Notte}
\author[Zeissaddress]{J. Huang}

\author[hzdraddress]{S. Facsko}

\address[hzdraddress]{Helmholtz--Zentrum Dresden--Rossendorf e.V., Bautzner Landstr. 400, 01328 Dresden, Germany}
\address[tudaddress]{Technical University Dresden, 01062 Dresden, Germany}
\address[Zeissaddress]{Ion Microscopy Innovation Center at Carl Zeiss
	 Microscopy LLC, One Corporation Way, Peabody, Massachusetts 01960 USA}

\begin{abstract}
Time of flight backscattering spectrometry (ToF--BS) was successfully implemented in a helium ion microscope (HIM). Its integration introduces the ability to perform laterally resolved elemental analysis as well as elemental depth profiling on the nm scale. A lateral resolution of $\leq$ 54\,nm and a time resolution of $\Delta t \leq$ 17\,ns $(\Delta t/t \leq 5.4\%)$ are achieved. By using the energy of the backscattered particles for contrast generation, we introduce a new imaging method to the HIM allowing direct elemental mapping as well as local spectrometry. In addition laterally resolved time of flight secondary ion mass spectrometry (ToF--SIMS) can be performed with the same setup. Time of flight is implemented by pulsing the primary ion beam. This is achieved in a cost effective and minimal invasive way that does not influence the high resolution capabilities of the microscope when operating in standard secondary electron (SE) imaging mode. This technique can thus be easily adapted to existing devices. The particular implementation of ToF--BS and ToF--SIMS techniques are described, results are presented and advantages, difficulties and limitations of this new techniques are discussed.

\end{abstract}


\end{frontmatter}


\section{Introduction}

In the recent past helium ion microscopy~\cite{Hlawacek2013a} has become a mature technique that is best known for its high resolution imaging capabilities. The latest version of these devices, the Zeiss helium ion microscope (model \textit{Orion NanoFab}) (used in this work) is able to operate with He as well as with Ne ions and provides high resolution nano--engineering capabilities~\cite{Alkemade2012b, Kuznetsov2014, Bell2009}, that so far are unmatched by any other technique. Using neon in the gas field ion source (GFIS) nano--structuring with 2\,nm lateral resolution is possible without any metal (Ga) contamination~\cite{Cybart2014,Notte2012}. Although exceptional nano machining and imaging results on insulating and biological samples have been achieved, so far no analytical elemental information can be obtained in the HIM. 

Several attempts have been made in the past to obtain analytical information utilizing the nano--sized ion beam available in GFIS microscopes. Early attempts to perform backscattering spectrometry (BS) in a HIM utilized a cooled, windowless silicon drift detector~\cite{Sijbrandij2008}. However, Si particle detectors provide an energy resolution with a low $\Delta E/E$ ratio of just 1:10 or worse. The so obtained BS spectra have been useful only in a limited number of specialized cases. Further, from this attempt it became clear that monolayer sensitivity should in principle be possible~\cite{Sijbrandij2010}. Analyzing the energy distribution of the emitted secondary electrons for elemental analysis has not matured so far~\cite{Petrov2011}. Here, matrix effects and non--linearities in the SE--yield hinder the quantification of the obtained SE energy spectra~\cite{Ramachandra2009a,Joy2011}. Recently impressive progress has been made in the development of a dedicated SIMS add--on for the HIM~\cite{Pillatsch2013, Wirtz2012}. The approach followed by Wirtz \textit{et al.}~\cite{Wirtz2015} will allow high resolution SIMS spectra and mass filtered images with sub-20\,nm lateral resolution.

For the elemental analysis by BS several different approaches could be used. For conventional primary ion energies in the range from 100\,keV to some MeV various approaches of backscattering energy measurement have been established in the past. Semiconductor detectors are most commonly used and can deliver an energy resolution down to 5.1\,keV for 2.25\,MeV protons~\cite{Klingner2013} using an in--vacuum preamplifier. Using additional detector cooling an energy resolution of 1.8\,keV was reported for 600\,keV deuterons~\cite{Primetzhofer2011b} and 7\,keV for 3.2\,MeV~He~\cite{Steinbauer1994a}. For low energies such as the ones used in HIM (typically 10\,keV to 40\,keV) energy resolution of 4.5\,keV for 25\,keV He particles have been reported. These results have been achieved by using a Peltier cooled silicon drift detector~\cite{Sijbrandij2008,Sijbrandij2010}. Other approaches make use of magnetic~\cite{Kimura2004} or electrostatic energy analyzers which have an excellent energy resolution down to $\Delta E/E \leq 0.001$ but are only sensitive to charged particles \cite{TerVeen2009, Brongersma2010} and acquire spectra in a sequential manner.

The fraction of charged, backscattered projectiles for energies below 10\,keV decreases rapidly with increasing depth and is below one percent for scattering from depths as low as one nm~\cite{Buck1975, Draxler2002, Primetzhofer2008a, Primetzhofer2011e}. For energies above 30\,keV the charge fraction stays below ten percent~\cite{Buck1973}. 

Consequently, for backscattered particle detectors that are sensitive only to charged particles, the overall usefulness is reduced due to the increased sample damage and longer analysis time. The attempt of performing BS in a HIM is connected to a very small beam size and low primary ion energies. It thus is clear that in order to prevent sample damage (by sputtering and/or bubble formation~\cite{Wilson1976, Livengood2009, Tan2010, Veligura2013a}) backscattered particle detection has to be sensitive to both backscattered ions as well as neutrals.

Micro calorimeters would provide the necessary energy resolution~\cite{Wollman1997} and are sensitive to ions and neutrals but their implementation into the microscope and the decoupling from the heat reservoir of the chamber would require a considerable amount of investigation and engineering work. 

The most convenient approach is the application of ToF spectrometry. Performing ToF spectrometry by triggering the start signal from secondary electrons from the sample surface are currently under development for classical Ga focused ion beams~\cite{Abo2012} as well HIM~\cite{Kobayashi2014a, Xu2014}. However, the high number of emitted SEs compared to the rather low cross sections for backscattering lead to a very low coincidence rate and subsequently a poor signal to noise ratio and therefore long measuring times.

Different to previous approaches, here time of flight spectrometry is enabled by pulsing the primary ion beam. We present first analytical results obtained with a combined time of flight backscattering spectrometry and secondary ion mass spectrometry setup. Both techniques utilize the same cost efficient approach, which requires minimal modifications of the system to ensure that the high resolution imaging capabilities are maintained when no analytical information is required. Switching between ToF--BS and standard SE imaging can be performed electronically and requires no mechanical adjustments on the instrument hardware. 

\section{Experimental}

\begin{figure}[hbt]
	\includegraphics[width=\linewidth]{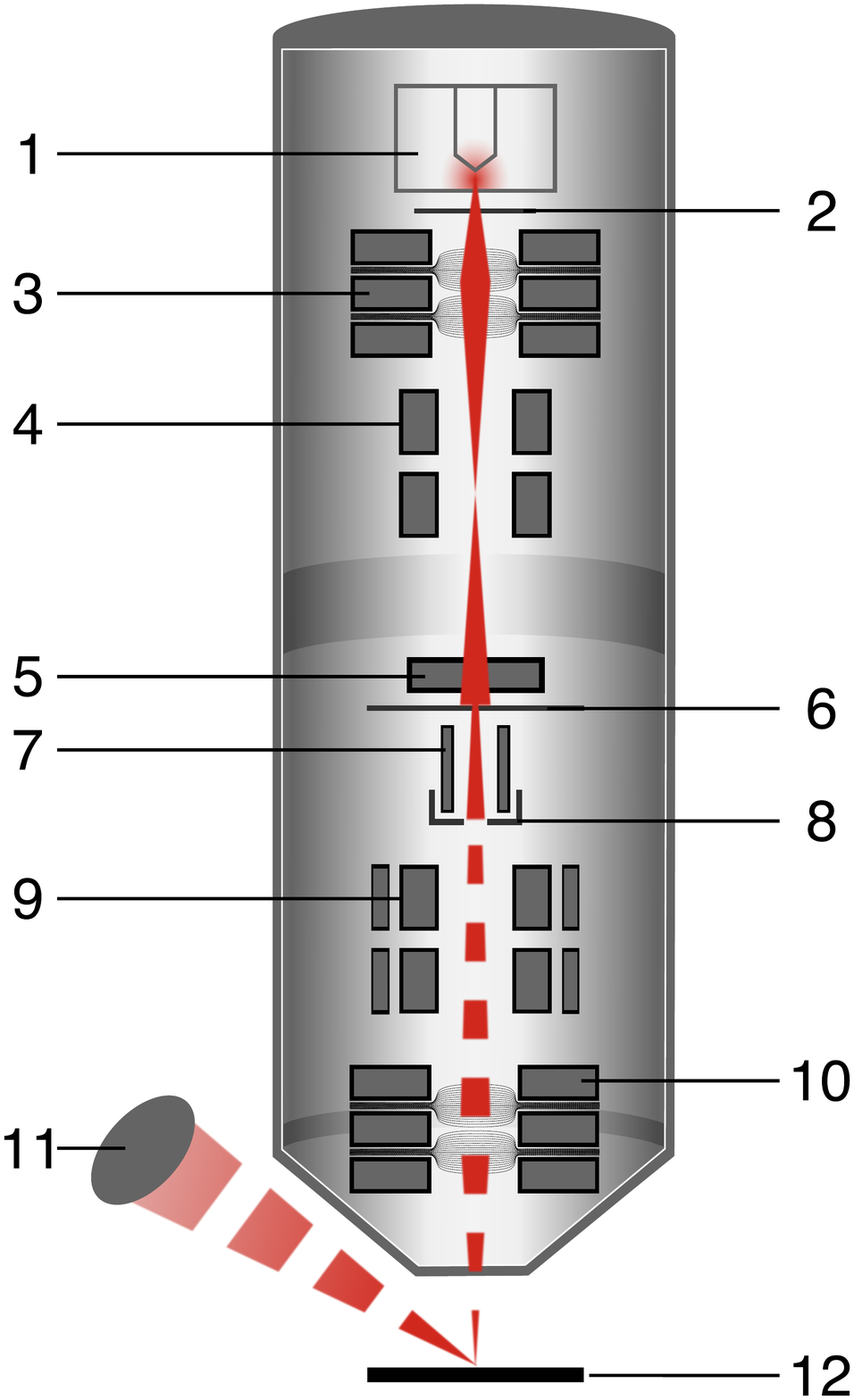}
    \caption{\label{scheme_him} Simplified scheme of the HIM (not to scale): 1~Source and gas chamber, 2~Extractor, 3~Einzel lens I, 4~Quadrupole, 5~Column isolation valve,	 6~Aperture, 7~Blanking unit, 8~Faraday cup, 9~Octopole, 10~Einzel lens	 II, 11~Micro channel plate, 12~Sample. The beam path is indicated in red.} 
\end{figure}

The helium ion microscope delivers primary ion energies from 5\,keV to 35\,keV, typical ion currents of a few pA and a beam focus below 0.5\,nm. Higher currents of up to 150\,pA are possible, however only with a larger beam spot and consequently a lower lateral resolution. A scheme showing the major components of the device is presented in fig.~\ref{scheme_him}.

\begin{figure}[hbt]
	\includegraphics[width=\linewidth]{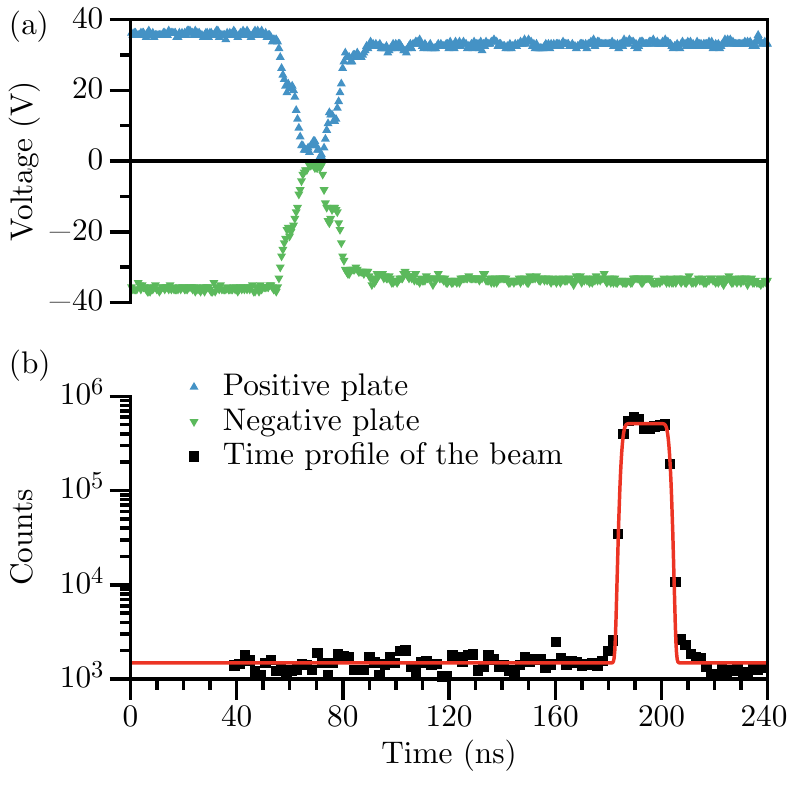}
    \caption{\label{pulse_timing} Voltage pulses on both blanking plates (blue and blue triangles (a)) and time profile of a pulsed 30\,keV~He$^+$ ion beam (black squares in (b)), both triggered by a TTL pulse at 0\,ns. The beam pulse fits a box profile with a FWHM of 17\,ns and a rise/fall time of 1.7\,ns (red line). It starts at 120\,ns after unblanking the ion beam which approximately corresponds to the flight time from the blanker to the sample.} 
\end{figure}

The start signal for the ToF measurement is created by pulsing the primary ion beam. To retain the excellent imaging capabilities of the microscope no changes have been made to the ion beam column. A newly designed fast pulsing electronics has been added to the column--mounted electronics of the beam blanking unit ((7) in fig.~\ref{scheme_him}). The new electronics generates fast voltage pulses on both blanking plates that unblank the ion beam from the Faraday cup ((8) in fig.~\ref{scheme_him}) for a few nanoseconds towards the sample. It is triggered by a standard TTL pulse from a pulse generator with a typical repetition rate of up to 500\,kHz. An oscillograph of the voltages on both blanking plates is shown in fig.~\ref{pulse_timing}(a). A rise/fall time of 8\,ns equally for both blanking plates was achieved. 

The stop signal for our ToF measurements is obtained by detecting the backscattered particles on a micro channel plate (MCP) referred to as the stop detector in the following ((11) in fig.~\ref{scheme_him}). It is a chevron stack MCP (model AF2225-A41D, type F1217-01 Hamamatsu Photonics) operated at an amplification voltage of 1800\,V. The stop detector is mounted under a backscattering angle of 126$\degree$ to the primary ion beam and in a distance of 358\,mm to the target surface with a solid angle of 10.8\,msr. For an increased relative time resolution a second MCP is mounted in a distance of 1023\,mm which results in a smaller solid angle of 1.3\,msr. The stop signal is amplified by a pre--amplifier (model TA2000B--2, FAST ComTec), the edge detection is done by a constant fraction discriminator (model 2128, FAST ComTec) and the time of flight is measured with a time to amplitude converter (model 2145, Canberra) and digitised by an analog to digital converter (model 7072T, FAST ComTec). Standard spectroscopic equipment (pulse height analysis via a multi channel analyzer) finally reveals the ToF spectrum.
   
The performance of the ToF setup has been evaluated by direct measurement of the time profile of the pulsed ion beam using a channeltron mounted on the sample stage. The time profile of a 30\,keV pulsed He ion beam has been integrated over 2 $\times$ 10$^7$ pulses and is shown in fig.~\ref{pulse_timing}(b). It can be described by a double error function with a width of 17\,ns and a rise/fall time of 1.7\,ns. 

\section{Results and discussion}

\subsection{Time of flight backscattering spectrometry}

\begin{figure}[hbt]
 \includegraphics[width=\linewidth]{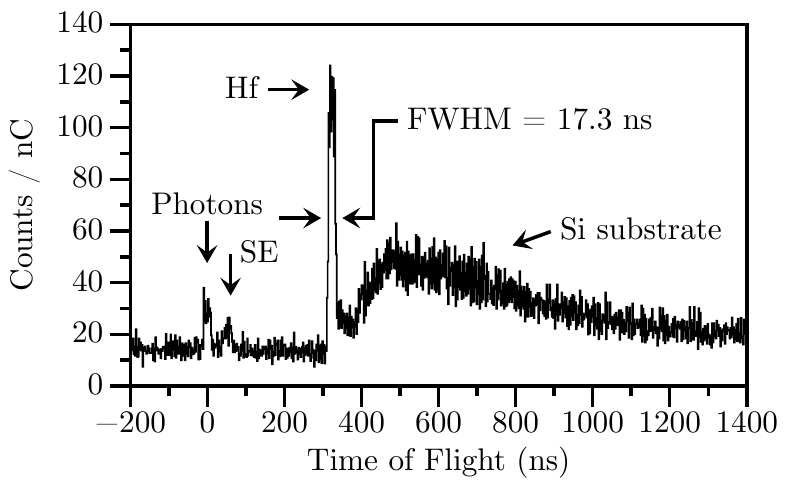}
 \caption{\label{ToFHfO2} Time of flight spectrum of 30\,keV~He$^+$ backscattered from a 2\,nm HfO$_2$ layer on top of Si measured with 17.3\,ns time resolution. The time scale of the spectra was calibrated by helium induced photons. The total charge collected for this spectrum was 1.9\,nC.} 
 \end{figure}

A typical ToF He backscattering spectrum of a 2\,nm~HfO$_2$ layer on top of Si is shown in fig.~\ref{ToFHfO2}. For this measurement the pulsed beam was continuously scanned across a sample area of 200\,\textmu m$^2$. The peak at 320\,ns corresponds to backscattering from Hf which is separated from the signal of the silicon bulk material starting at 380\,ns. Since the HfO$_2$ layer is very thin, its full width half maximum (FWHM) corresponds to the time resolution of the ToF setup. The measured $\Delta t = 17.3\,\rm{ns}$ equals a relative time resolution of $\Delta t / t \leq$ 5.4\,\%. This value fits to the ratio between the length of the blanking plates and the distance between sample surface and stop detector (5.8\%). ~
Obviously, increased energy straggling inside the sample with increasing depths~\cite{VanGastel2015} has to be taken into account and conventional single collision analysis can be assumed only for near-surface scattering. ToF spectrometers also allow even better energy resolution by increasing the flight path at cost of decreased solid angle and counting statistics. Using the same setup but with a flight path of 1023\,mm and a flight time of 900\,ns, a relative time resolution of $\leq$ 2.7\% has been achieved. ~

The layer structure of a sample and the elemental composition of the particular layers can be determined from the measured spectra by simulation and comparison of the simulation result to the data in an interactive way. Conventional backscattering spectrometry is typically performed with primary energies above 1\,MeV and well known as Rutherford backscattering spectrometry (RBS). Analytical simulation software packages like RUMP~\cite{Doolittle1985}, WINDF~\cite{Barradas1997} or SIMNRA~\cite{Mayer1999} are commonly used and known to deliver quite accurate results. All of them have in common that they assume a single or at most two main collision (besides simple models to correct the effect of multiple scattering) leading to a change of the direction and the energy of the primary particle. However, in the low energy range below 100\,keV the majority of backscattered particles are suffering multiple large angle collisions with the target atoms. Thus these programs fail to recover the measured spectra. In contrast Monte Carlo simulation software like SRIM~\cite{Ziegler2008}, TRIDYN~\cite{Moller1984}, CORTEO~\cite{Schiettekatte2008} or TRBS~\cite{Biersack1991} use a binary collision approximation and deliver results taking into account multiple scattering. 

The comparison of spectra from ToF measurements with simulated spectra requires a conversion of the time of flight into an energy or vice versa. A precise knowledge of the offset of the time axis is therefore essential. Since the start signal is triggered by blanking the beam and the stop signal by the backscattered particle hitting the stop MCP, the measured time of flight has to be reduced by the electronic delay and the flight time of the primary ions from the blanker to the sample surface. The latter one depends on the ions mass, its energy, and the distance from the blanking plates to the sample and can be embedded in the analysis routine. However, the first part is more difficult. Therefore the total time offset is calibrated by making use of photons emitted during the interaction of the primary beam with the sample. The lifetime of the excited states of $\leq 10$\,ns~\cite{Andersson1998, Hagstrum1976} and the ToF below 2\,ns make them suitable for the calibration. Although the production rate for photons in this ion energy range is rather small it is sufficient to collect a usable signal in reasonable time (minutes). 

\begin{figure}[hbt]
	\includegraphics[width=\linewidth]{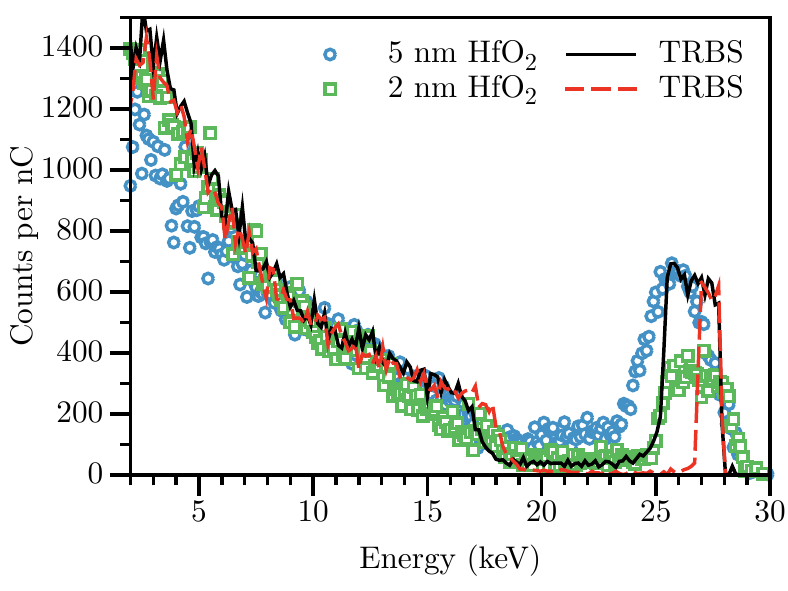}
	\caption{\label{EnergyHfO2}ToF spectra of 5\,nm and 2\,nm~HfO$_2$ layer on top of Si converted into energy space (dots and squares) and corresponding TRBS~\cite{Biersack1991} simulations (black and red lines). The collected charge has been adapted to fit the Si bulk signal to the TRBS simulation.}
\end{figure}

The converted spectrum of the ToF measurement presented in fig.~\ref{ToFHfO2} (30\,keV~He on 2\,nm HfO$_2$ layer on Si) is plotted in fig.~\ref{EnergyHfO2} together with a spectrum of a 5\,nm HfO$_2$ layer on Si. The dots and squares present measured spectra while the black and red lines are results from Monte Carlo simulations using TRBS. The energy resolution of our ToF setup translates to a sufficient depth resolution to clearly distinguish between the different thicknesses of the HfO$_2$ layers. In each simulation the trajectories of 5 $\times 10^7$ He ions were evaluated. Simulated particles backscattered towards the detector were recorded and sorted into a pulse height spectrum according to their energy. The resulting spectrum was scaled to the solid angle and plotted as counts per nC. 

The collected charge in both measurements has been adapted to fit the Si bulk signal to the TRBS simulation. The charge adaption had to be done because the charge measurement in the HIM is designed to measure DC ion currents instead of a pulsed beam. For both measurements two different scaling factors had to be applied since they were recorded with different primary ion beam currents. For both samples the gap between Hf-peak and Si substrate reveals a non-zero offset that is not predicted by the simulations. A similar observation can be found for focused ion beam based ToF--BS~\cite{Hayashi2004, Abo2012}. The origin of this effect stays unclear so far. The larger height and the smaller width of the simulated peaks originate from neglecting any detector resolution which is present in the experimental data only.

\subsection{Elemental mapping with backscattering spectrometry}

To obtain laterally resolved element maps we made use of a self-made micro controller based external scan electronic. This external scan controller provides analog signals to the input of the microscope steering the scanning of the beam. It further records the time of flight for each event from the analog to digital converter together with the current scan position. These events are stored in a list mode file for further evaluation. Thus one can post-select particular regions of interest within the scan field and extract local energy spectra. Scan parameters like field of view, number of pixels or dwell time (pixel time) are configurable in the data acquisition software.

\begin{figure}[hbt]
 \includegraphics[width=\linewidth]{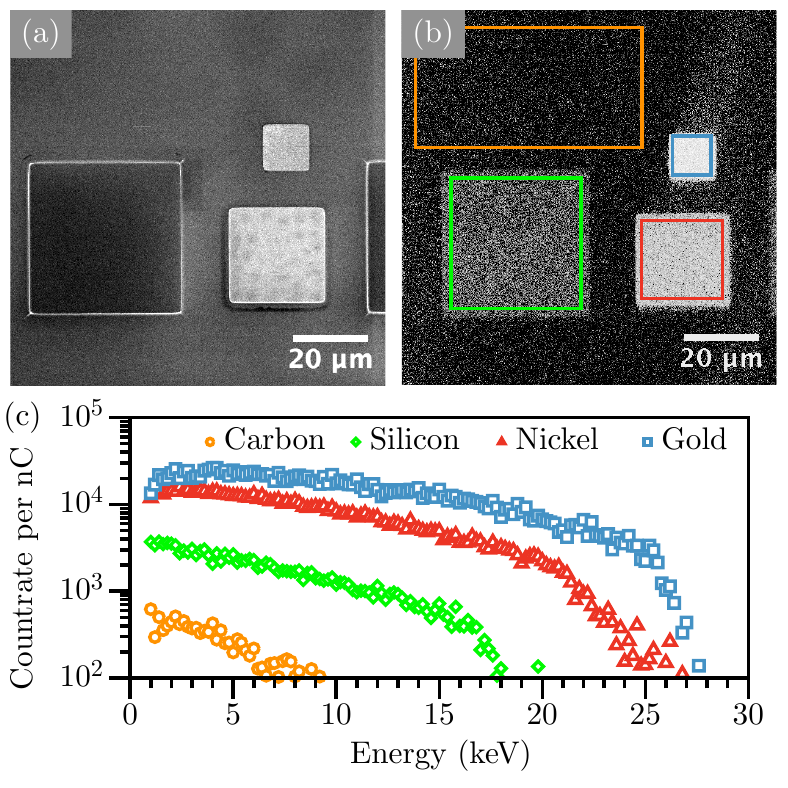}
 \caption{ \label{imaging_modes}Images of a Au/Ni/Si patterned test sample (as described in the text) acquired in SE mode (a) and in ToF--BS mode (b). (c) shows energy spectra of backscattered He within different regions of interest in (b). The color of each spectrum in (c) corresponds to the colors of the rectangles marking the corresponding region of interest in (b).}
\end{figure}

To test the imaging capabilities of the HIM in ToF--BS mode we used a glassy carbon sample coated with rectangular patches of Si, Ni and Au. The patches have different dimensions of 40\,\textmu m $\times$ 40\,\textmu m $\times$ 300\,nm (Si), 25\,\textmu m $\times$ 25\,\textmu m $\times$ 110\,nm (Ni) and 12\,\textmu m $\times$ 12\,\textmu m $\times$ 85\,nm (Au), respectively. An image of the test sample in standard SE mode is shown in fig.~\ref{imaging_modes}(a) and the ToF--BS image from the same surface region is presented in fig.~\ref{imaging_modes}(b). For this image only the highest backscattering energy in each pixel is taken for contrast generation. This leads to an enhanced elemental contrast. In fig.~\ref{imaging_modes}(c), the BS spectra obtained from different regions within the image presented in fig.~\ref{imaging_modes}(b) are shown. This allows local quantitative element analysis which is currently not possible in standard SE imaging. 

Partially blanked ions lead to non axial trajectories and a spatial offset. The flight time of a 30\,keV He ion through the blanking plates is approximately 17\,ns. The ion will pass the blanker in an undisturbed manner if the plates are grounded during its transition. However, if the blanker changes state during the transition of the ion, it will be deflected from the aligned path through the column. This leads to a reduced lateral resolution in pulsed beam operation. The lateral resolution parallel to the deflection direction is most influenced by this effect. 

\begin{figure}[hbt]
 \includegraphics[width=\linewidth]{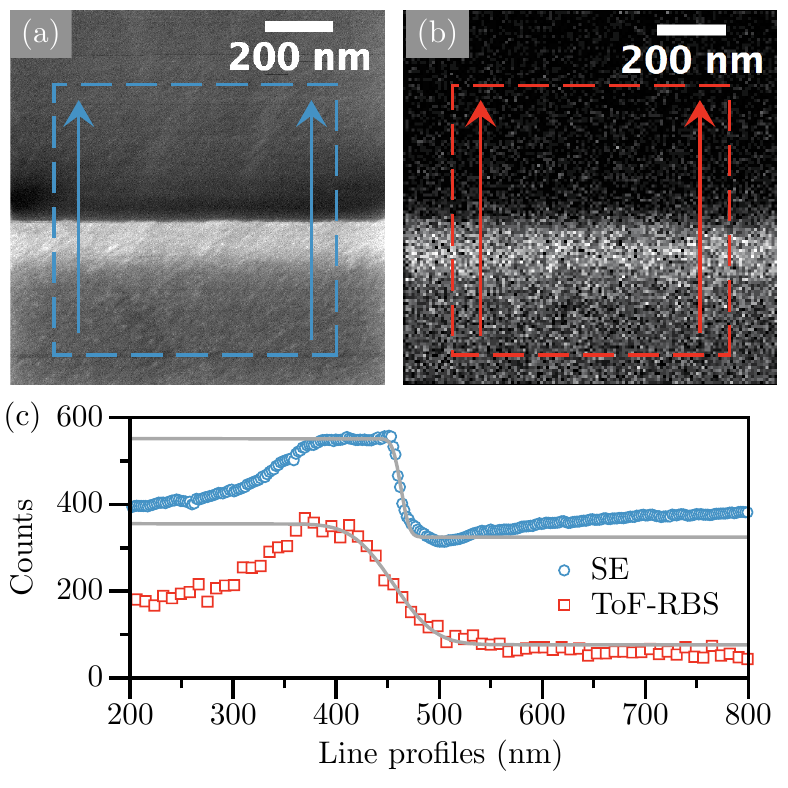}
 \caption{ \label{edge_resolution}SE image without pulsing the ion beam (a) and ToF--BS image (b) of a Ni patch on the test sample described in the text and derived line profiles of the Ni edge (c). Line profiles are measured and averaged across the Ni edge according to the rectangles plotted in (a) and (b) and error functions fitted at the edges. The edge resolutions (80\%-20\%) was determined to 10.9\,nm in SE mode and 53.7\,nm in ToF--BS mode using 55\,ns beam pulses.}
\end{figure}

The edge resolution in pulsed mode was evaluated using a Ni patch on our test sample. The results are shown in fig.~\ref{edge_resolution}. Images of the Ni patch without pulsing the ion beam in SE mode (fig.~\ref{edge_resolution}(a)) and in ToF--BS mode (fig.~\ref{edge_resolution}(b)) as well as the corresponding line profiles across the edges (fig.~\ref{edge_resolution}(c)) are presented. Line profiles of several neighboring (vertical) lines were averaged (indicated by the rectangles in fig.~\ref{edge_resolution}(a,b)) leading to a better signal to noise ratio. The blanking direction in this measurement was 52$\degree$ with respect to the Ni edge. The edge resolution (80\%-20\%) is 10.9\,nm in SE mode without pulsing the beam and 53.7\,nm in ToF--BS mode using beam pulses with a length of 55\,ns. 

The reduced lateral resolution in ToF--BS mode is attributed to the partial blanking of the beam as discussed above. The larger sampling volume of the backscattered particles and the sample drift due to longer acquisition time contribute further to the reduced lateral resolution.

It should be mentioned that the pulse length influences lateral resolution, energy resolution and signal to noise ratio simultaneously. By adjusting pulse length and duty cycle one can vary between optimum lateral and highest energy resolution. Both have to be adapted according to the particular demands of the measurement task.

\subsection{Time of flight secondary ion mass spectrometry}

In addition to the possibility of the measurement of ToF--BS spectra our approach of pulsing the primary ion beam allows time of flight secondary ion mass spectrometry (ToF--SIMS). For this purpose the sputtered ions (with energies of few to few tens eV~\cite{Dowsett2012}) have to be accelerated to higher kinetic energies and guided towards the MCP. The time of flight directly scales with the secondary ion mass in this mode of operation. 

\begin{figure}[hbt]
 \includegraphics[width=\linewidth]{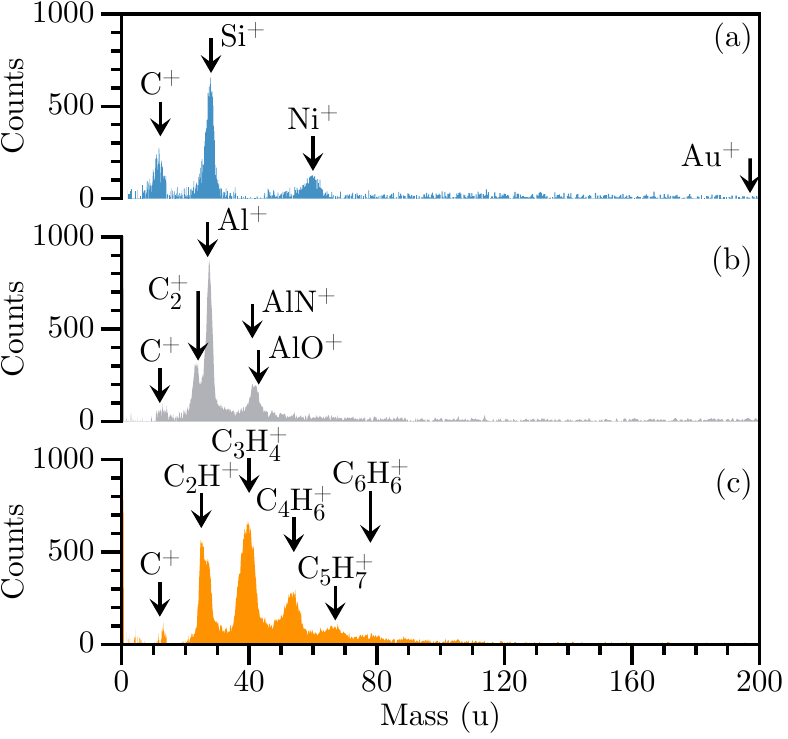}
 \caption{ \label{ToF_sims}ToF-SIMS spectra of the Au/Ni/Si/C test sample describe in the text (a), an aluminium sample (b) and copper tape (c) measured in the HIM with 25\,keV Ne ions and 250\,ns pulse length.}
\end{figure}

In our experiments we applied an acceleration voltage of 500\,V leading to an additional energy of 500\,eV for single charged and 1000\,eV for doubly charged particles, respectively (both can be identified in the mass spectra). To minimize the time spread caused by different starting energies, acceleration has to be applied as close as possible to the surface. Therefore we biased the sample holder to 500\,V and mounted a grounded TEM grid on top of the sample in a distance of less than one millimeter. The sample was additionally tilted to face towards the MCP. Therefore the sputtered ions can pass through the chamber to the MCP without the need of a flight tube. Due to the sample bias, the primary ion beam is decelerated from 30\,keV to 29.5\,keV before reaching the sample. This setup enables SIMS measurements with moderate efforts, but with inferior ion collection efficiency compared to dedicated SIMS machines were ions are extracted by more complex extraction optics and guided towards the detector. Due to the finite mesh size of the TEM grid inhomogeneities of the electric field may occur leading to a broadening of the beam focus. It has been shown by Dowsett \textit{et al.}~\cite{Dowsett2012} that an advanced extraction system would improve the efficiency of the SIMS setup in the HIM while keeping the lateral resolution below 10\,nm. Since the majority of emitted secondary electrons have energies less than 500\,eV~\cite{Ramachandra2009a, Petrov2010, Petrov2011} standard SE imaging is not available during ToF--SIMS measurements. Because of higher sputter yields the use of neon is preferred for SIMS experiments. 

The ToF--SIMS spectrum of the test sample described above is shown in fig.~\ref{ToF_sims}(a). Mass peaks from carbon, silicon and nickel are found well separated from each other. The gold peak cannot be distinguished from background noise because the secondary Au$^+$ yield is several orders of magnitudes smaller compared to carbon, silicon and nickel~\cite{Pillatsch2013}. In fig.~\ref{ToF_sims}(b,c) further ToF--SIMS spectra of a pure aluminum sample and a piece of copper tape are presented. For the latter the ToF--SIMS spectrum actually reveals the constituents of the organic glue on top of the copper which is much thicker than the origin of the sputtered particles.

Since the flight times of the accelerated sputtered ions ($E=$~500\,eV) are higher than those of backscattered He in ToF--BS mode ($E\leq 30\,\rm{keV}$) ToF--SIMS spectra can be acquired by using larger puls lengths. We used 250\,ns pulses corresponding to $t / \Delta t \approx 40$ and measured a FWHM mass resolving power of $M / \Delta M \approx 12$ (Al peak). The initial energy distribution~\cite{Dowsett2012} of sputtered particles leads to variations in the time of flight from the sample surface to the acceleration grid. This effect can be reduced by applying higher acceleration voltages. However, compared to the short pulses for ToF--BS measurements, higher pulse lengths lead to a better lateral resolution (see text above). The need for ion extraction by biasing the sample also contributes to a lateral spread (see also~\cite{Dowsett2012}). 

\begin{figure}[hbt]
 \includegraphics[width=\linewidth]{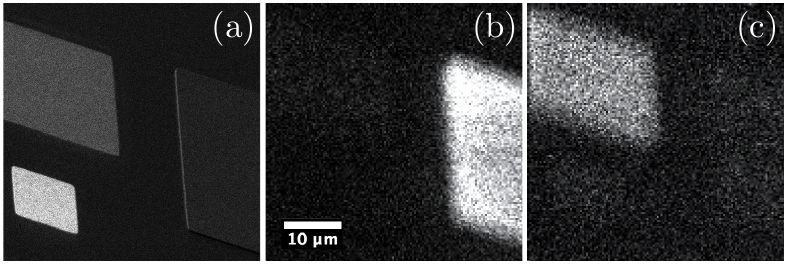}
 \caption{ \label{ToF_sims_2}Images of the carbon test sample described in the text acquired in SE mode (a) and in ToF--SIMS mode (b,c) using 25\,keV neon ion beam with 250\,ns pulses. Different mass filters were applied for generating maps of silicon (b) and nickel (c) distribution out of the common list mode file.}
\end{figure}

The SE image of the carbon test sample is shown in fig.~\ref{ToF_sims_2}(a) and ToF--SIMS images obtained from the same location are presented in fig.~\ref{ToF_sims_2}(b,c). For generating the particular ToF--SIMS images from the corresponding list mode file different mass (flight time) filters were applied. In fig.~\ref{ToF_sims_2}(b) only silicon is shown whereas in fig.~\ref{ToF_sims_2}(c) the nickel counts are presented. Since in ToF--SIMS mode the sample has to face towards the MCP detector (see text above) the rectangular patches appear as parallelogram shapes in the images. Although mass filters are applied in fig.~\ref{ToF_sims_2}(b,c) one can identify faintly visible structures at the position of the remaining patches. These originate from neutrals (mainly backscattered neon) which could be suppressed by a reflectron flight tube in a future design. However, characterization and optimization of the ToF--SIMS mode with respect to mass and lateral resolution as well as signal--to--noise ratio are subject of future investigations. As is evident from the presented data ToF--SIMS in the HIM is perfectly capable of delivering an excellent elemental contrast for imaging purposes. However, quantification of elements in mixed layers can not be done from pure SIMS measurements without comparison to standards. This drawback of SIMS is partly overcome here as our setup is capable to also measure ToF--BS spectra. These deliver the needed quantitative information on the layer composition. Thus ToF--BS and ToF--SIMS performed in--situ complement each other and therefore deliver a maximum of compositional information on the sample.

\section{Summary}

We demonstrated that time of flight backscattering spectrometry as well as secondary ion mass spectrometry can be performed in a helium ion microscope to obtain information on the elemental composition of a sample. This information is not accessible in standard SE imaging mode. Data acquisition in list mode enables post--processing of measured data to obtain BS spectra on specific regions of interest and elemental mapping at the nanometer scale. A lateral resolution of 54\,nm for ToF--BS imaging was demonstrated. Spatial resolved BS was so far only possible down to 300\,nm$^2$~\cite{Reinert2006, Watt2003} using ion micro probe experiments requiring big (MeV) ion accelerators. Our experimental approach requires a minimum of changes to the existing HIM hardware and thus may be easily retrofitted on existing devices significantly enhancing their capabilities. The setup additionally allows ToF--SIMS measurements in the HIM delivering excellent elemental contrast. In summary we present a minimal invasive and cost effective way to extract a maximum of information from the sample in a correlative approach. The ability to obtain SE, ToF--BS and TOF--SIMS images in--situ, enables the user to correlate these data and in this way obtain elemental and topographical information at the nanometer scale.

\section*{Acknowledgement}

Financial support from the Bundesministerium f\"ur Wirtschaft und Energie (BMWi) (Grant 03ET7016) is acknowledged. The authors thank R.~Aniol (HZDR) for manufacturing of the mechanical parts for the ToF setup and P.~Bauer (JKU Llinz) for providing the TRBS simulations.

\bibliographystyle{elsarticle-num}
\bibliography{library}

\end{document}